\documentclass[preprint,prl,aps,floats,amsfonts,amssymb,stmaryrd,showpacs]{revtex4}
\usepackage{amsmath}
\usepackage[dvips]{graphicx}
\usepackage{natbib}

\def\EQ{\begin{equation}}
\def\EN{\end{equation}}
\def\EQA{\begin{eqnarray}}
\def\ENA{\end{eqnarray}}
\def\vv{{\bf v}}

\def\VV{{\bf V}}
\def\bb{{\bf b}}
\def\BB{{\bf B}}

\def\A{\Omega}

\def\T{\mathcal{T}}

\begin{document}

\title{Dynamo quenching due to shear flow}
\author{Nicolas Leprovost and Eun-jin Kim}
\affiliation{Department of Applied Mathematics, University of Sheffield, Sheffield S3 7RH, UK}

\begin{abstract}
We provide a theory of dynamo ($\alpha$ effect) and momentum transport in three-dimensional magnetohydrodynamics. For the first time, we show that the $\alpha$ effect is severely reduced by flow shear even in the absence of magnetic field. The $\alpha$ effect is further suppressed by magnetic fields well below equipartition (with the large-scale flow) with different scalings depending on the relative strength of shear and magnetic field. The turbulent viscosity is also found to be significantly reduced by shear and magnetic fields, with positive value. These results highlight a crucial effect of shear and magnetic field on dynamo quenching and momentum transport reduction, with important implications for astrophysical and laboratory plasmas, in particular for the dynamics of the Sun.
\end{abstract}


\maketitle

\noindent{\it Introduction.}- Dynamo action describes the process through which a motion of a conducting fluid in the presence of a magnetic field amplifies that magnetic field. This is a fundamental mechanism that explains ubiquitous magnetic fields in a variety of systems, including astrophysical, geophysical and laboratory plasmas \cite{Parker55, Moffatt78,Guzdar02}. This is especially the case in stellar convective regions where the diffusive time-scale is so short due to turbulent motion that any primordial field would decay in a few million years. In a conducting fluid of velocity ${\bf V}$,  magnetic field ${\bf B}$ evolution is governed by the induction equation:
\EQ
\partial_t {\bf B} + {\bf V} \cdot {\bf \nabla} \BB =  \BB \cdot {\bf \nabla} \, {\bf V} + \eta \nabla^2 {\bf B} \quad  \text{and} \quad {\bf \nabla \cdot \BB} = 0 \; , \label{Induction}
\EN
where $\eta$ is the ohmic diffusivity. In case of the Sun, the prominent radial shear layer (tachocline) permits the creation of a toroidal magnetic field from an existing poloidal one via shearing [the third term in Eq. (\ref{Induction}), so-called $\Omega$ effect]. The generation of poloidal field from the toroidal field requires  another mechanism such as $\alpha$ effect via kinetic helicity, magnetic buoyancy \citep{Cline03} or some kind of magnetic instability \citep{Rincon07} and has proved to be more difficult as explained below. 

In the presence of turbulence, the magnetic and velocity fields can be decomposed into a mean and fluctuating parts: ${\bf B} = \langle {\bf B} \rangle + {\bf b}$  and ${\bf V} = \langle {\bf V} \rangle + {\bf v}$, where the $\langle \bullet \rangle$ stands for an average on the realization of the small-scale fields. Substitution of this decomposition into Eq. (\ref{Induction}) and averaging yield the following equation for the mean magnetic field:
\EQ
\label{Induction2}
\partial_t \langle {\bf B} \rangle + \langle {\bf V} \rangle \cdot {\bf \nabla} \langle \BB \rangle =  \langle \BB \rangle \cdot {\bf \nabla} \, \langle {\bf V} \rangle + \eta \nabla^2 \langle {\bf B} \rangle + {\bf \nabla} \times {\bf \epsilon} \; ,
\EN
where ${\bf \epsilon} = \langle {\bf v} \times {\bf b} \rangle$ is the electromotive force. In the framework of mean-field dynamo theory \citep{Moffatt78} ${\bf \epsilon}$ is linear in the mean magnetic field with the following expansion:
\EQ
\label{Electromotive}
\epsilon_i = \alpha_{ij} \langle B_j \rangle + \beta_{ijk} \nabla_j \langle B_k \rangle + \dots  \; .
\EN
where $\alpha$ and $\beta$ are tensors. The symmetric part of  $\alpha$ acts as a source term in  Eq. (\ref{Induction2}), creating poloidal (resp. toroidal) field from toroidal (resp. poloidal) one. The antisymmetric part of the $\alpha$ tensor is interpreted as a mean pumping. The symmetric part of the $\alpha$ effect has been thought to be generated by a helical turbulence, which is likely to be induced by Coriolis force in stellar convection zones. This type of dynamo is thus classified as $\alpha\Omega$ if the $\Omega$ effect is stronger than this effect, or $\alpha^2$ type if this effect dominates over the $\Omega$. In convecting stars without pronounced differential rotation, the latter is the only possible mechanism as there is no $\Omega$ effect. The antisymmetric part of $\beta$ is just a turbulent diffusivity which adds up to the molecular one whereas the symmetric part may contain additional source terms.

When the large-scale magnetic field is sufficiently strong, it begins to influence the velocity field. At some point it will severely quench the generation coefficient (such as the $\alpha$ effect), thereby saturating dynamo action. The effect of an uniform magnetic field on magnetohydrodynamical (MHD) turbulence was addressed by \cite{Rudiger74} who showed that the $\alpha$ effect is reduced both for weak and strong magnetic field. However, the effect of fluctuating magnetic fields on the $\alpha$ effect remains controversial. It has been argued that the dynamo saturates when $\langle B \rangle \propto R_m^{-n} \sqrt{\langle {\bf v}^2 \rangle}$ due to small-scale magnetic fields. Here, $R_m$ is the magnetic Reynolds number. If the small-scale magnetic field grows preferentially to equipartition \citep{Cattaneo96}, the large-scale magnetic field must be far below equipartition with the previous formula holding with $n=1$. Alternatively, if the growth of the small-scale field is limited \citep{Blackman96}, the large scale field can almost reach equipartition and the previous formula holds for $n=0$.

The purpose of this Letter is to investigate a novel saturation mechanism which has received little attention: the effect of a stable large-scale shear flow. On the one hand, strong shear is good for dynamo as it creates magnetic energy via the $\Omega$ effect. But, a strong shear can reduce turbulent transport via shear stabilization \cite{Burrell97}. Indeed, the shear has been shown to significantly reduce the turbulence intensity and the turbulent transport of angular momentum, particle mixing and magnetic diffusion \cite{Kim05,Kim06,2Shears}. As the shear might have a similar effect on the generation of magnetic field, it is crucial to compute self-consistently the $\alpha$ effect incorporating the effect of shear (or differential rotation). This is especially the case for solar dynamo which is often envisioned to take place at the base of the convection zone, where the shear is quite strong, e.g. to compensate the weakness of the interface dynamo \cite{Dikpati05}. Should the $\alpha$ effect be significantly quenched due to shear, it would put severe constraints on the magnetic fields that can be created by dynamo action. For instance, the dynamo number which characterizes the efficiency of the $\alpha\Omega$ dynamo is given by $D= \alpha \Omega L^3 / (\eta+\beta)^2$ \citep{Kulsrud99}, where $\Omega$ is the differential rotation and $L$ is a characteristic scale of the system. In this Letter, we show that the intensity of the $\alpha$ effect depends on the strength of the differential rotation, and consequently that the dynamo number is not simply proportional to the shear intensity. We also provide a consistent theory of momentum transport in sheared magnetized plasmas, necessary to understand large-scale shear flows (e.g. radial differential rotation). 

We consider 3D MHD forced turbulence in an incompressible conducting fluid governed by:
\EQA
\nonumber
\partial_t \VV + \VV \cdot {\bf \nabla} \VV &=& - {\bf \nabla} p + \BB \cdot {\bf \nabla} \BB  + \nu \Delta \VV + {\bf f} \; , \\ 
{\bf \nabla} \cdot \VV &=& 0 \; ,
\ENA
together with Eq (\ref{Induction}). Here, ${\bf B}$ is the magnetic field given in the unit of Alfv\'en speed, $p$ is the total (hydrodynamical + magnetic) pressure, and ${\bf f}$ is the small-scale forcing. To study the effect of shear flows and magnetic fields on small-scale turbulence, we prescribe a large-scale flow of the form $\langle \VV \rangle = - x \A {\bf \hat{y}}$ and a constant large-scale magnetic field $\langle \BB \rangle = B_0  {\bf \hat{y}}$ parallel with $\langle \VV \rangle$. Then, to solve the equations for the fluctuating velocity field, $\vv = \VV - \langle \VV \rangle$, and magnetic field, $\bb = \BB - \langle \BB \rangle$, we use the quasi-linear approximation assuming that the interaction between fluctuating fields is negligible compared to the interaction between large and small-scale fields. Note that this is well justified for weak turbulence due to shear flow \citep[e.g. see][]{Kim05}. As a general solution to small-scale equations is not analytically tractable, we consider the two limits of weak magnetic field, where the effect of the shear dominates that of the magnetic field, and of strong magnetic field, where the effect of the magnetic field dominates that of the shear. Specifically, the case $\gamma = B_0 k_y / \A \ll 1$ (weak magnetic field) and $\gamma \gg 1$ (strong magnetic  field) are considered in the following, where $k_y^{-1}$ is a characteristic scale of turbulent motions.

Our main interest in this Letter are the Reynolds stress and the electromotive force, which determine the growth/decay of the large-scale velocity field and the large-scale magnetic field, respectively. First, the Reynolds stress gives a turbulent viscosity $\langle u_x u_y \rangle = \nu^T \A$, which adds up to the molecular viscosity. Second, for an uniform magnetic field, the electromotive force (\ref{Electromotive}) reduces to the term proportional to $\alpha$ only. Furthermore, as the large-scale magnetic field is in the $y$ direction, the electromotive force reduces to $\epsilon_i = \alpha_{iy} B_y$. The source term for the large-scale magnetic field is given by the curl of the electromotive force, for instance, the source term for $\langle B_z \rangle$ is $\partial_x \epsilon_y - \partial_y \epsilon_x$. Assuming that the variation in the $x$-direction is much larger than in $y$ as is the case for the large-scale flow $\langle \VV \rangle (x)$ (for instance in the thin solar tachocline, the gradient in the radial direction is the dominant one), the main source of poloidal magnetic field is $\partial_x \epsilon_y$. Consequently, we focus here on the $y$ component of the electromotive force
\EQ
\label{BS2}
\epsilon_y = \langle v_z b_x - v_x b_z \rangle \; ,
\EN
and, in the following, we set $\alpha = \alpha_{yy}$ and refers to it as the alpha effect.

To calculate the correlation functions involved in the transports coefficients [see Eq. (\ref{BS2})], we consider an incompressible forcing which is spatially homogeneous and temporally short correlated with the correlation time $\tau_f$. Specifically, in Fourier space, the correlation of the forcing is given by:
\EQA
\nonumber
\langle \tilde{f}_i({\bf k_1},t_1) \tilde{f}_j({\bf k_2},t_2) \rangle &=& \tau_f \, (2\pi)^3 \delta({\bf k_1}+{\bf k_2}) \, \delta(t_1-t_2)  \times \\ \label{Forcing}
&& \qquad \phi_{ij}({\bf k_2}) \; ,
\ENA
where the tilde denotes a Fourier-transform with respect to the spatial variable. As noted previously, the $\alpha$ effect can be linked to the helicity of the turbulent flow. Consequently, we consider a forcing with both a non-helical part (with energy spectrum $E$) and a helical part (with helicity spectrum $H$) given by:
\EQ
\phi_{lm}({\bf k}) = E(k) \left(\delta_{lm} - \frac{k_l k_m}{k^2} \right) + i \epsilon_{lmp} k_p H(k) \; . 
\EN
It is important to note that in the absence of shear or magnetic field, the kinetic energy and helicity of the homogeneous flow ${\bf v}_0$ driven by the forcing become:
\EQA
e_0 &\equiv& \langle {\bf v}_0^2 \rangle = \frac{2 \tau_f}{(2\pi)^2} \int d k \frac{k_H^2}{\nu k^2} E(k) \; ,\\ \nonumber
h_0 &\equiv& \langle {\bf v}_0 \cdot {\bf \nabla \times v}_0 \rangle = \frac{4 \tau_f}{(2\pi)^2} \int d k \frac{k^2}{\nu } H(k)  \; .
\ENA

\noindent{\it Weak magnetic field.}-- To solve the equations for the small-scale fields in the weak magnetic field regime, we expand any field $\psi$ in powers of $\gamma= B_0 k_y / \A$, $\psi = \psi_{0} + \gamma \psi_{1} + \dots$ and solve order by order. The leading term and the first correction to the turbulent transport coefficient are obtained from the first four terms in the preceding expansion. The turbulent viscosity is found to be even in $B_0$ whereas the electromotive force is odd: $\nu^T = \nu^{T}_0 + \nu^{T}_2 + \dots$ and $\alpha = \alpha_{1} + \alpha_{3} + \dots$.

To investigate the effect of shear we focus on the limit of strong shear where the shear is stronger than the diffusion, characterized by the small parameter $\xi = \nu k_y^2 / \A \ll 1$. Note that this limit is relevant in astrophysical applications such as in the solar tachocline where the shearing rate is much larger than the diffusion rate for reasonable values of the parameters. In this limit ($\xi \ll 1$), the turbulent viscosity can be obtained, after a long algebra, as:
\EQA
\nonumber
\nu^T_0 &=& \frac{\tau_f}{2 (2\pi)^3 \A^2} \int d^3 k \frac{k_H^2 E(k)}{k_y^2} \times \\
\label{Visco0}
&& \quad \left[-\frac{k_y^2}{k^2} + \frac{k_z^2}{k_y^2} \T({\bf k}) \right] \sim \xi^2 \frac{e_0}{\nu k^2}  \; .
\ENA
Here, $k_H^2 = k_y^2 + k_z^2$ and $\T({\bf k}) = \pi \vert k_y \vert / 2 k_H - \arctan(k_x/k_H)$. Eq. (\ref{Visco0}) is the kinematic result ($B_0 = 0$) and shows that $\nu^T$ is strongly reduced by the shear with scaling $\A^{-2}$ (see also \citep{Kim05}). In the 2D case ($k_z=0$), $\nu^T$ is negative (inverse energy cascade), whereas, in 3D the second term in Eq. (\ref{Visco0}) dominates over the first one, making the turbulent viscosity positive. It is interesting to note that $\nu^T$ is proportional only to the energy part of the forcing, independent of the helical part. The first correction term due to the magnetic field to the kinematic result is obtained at second order in $\gamma$ as follows:
\EQA
\nonumber
\nu^T_2 &=& \frac{\tau_f B_0^2}{9 (2\pi)^3 \A^4} \int d^3 k k_H^2 E(k) \Gamma(2/3) \left( \frac{3}{2\xi}\right)^{2/3} \times \\ 
\label{Visco2}
&& \qquad \T({\bf k})^2 \left[1-\frac{k_z^2}{3 k_y^2} \right] \sim \gamma^2 \xi^{4/3} \frac{e_0}{\nu k^2}\; .
\ENA
Here, $\Gamma$ is the Gamma function. The correction (\ref{Visco2}) scales as $B_0^2 \A^{-10/3}$ and can be positive or negative depending on the values of the parameters. In 2D ($k_z = 0$), Eq. (\ref{Visco2}) is obviously positive, slowing down the inverse cascade. In comparison, in 3D, the second term in (\ref{Visco2}) dominates the first making $\nu^T_2$ negative. In both cases, the correction term (\ref{Visco2}) is always of the opposite sign to the leading order term (\ref{Visco0}). Comparing Eq. (\ref{Visco2}) and Eq. (\ref{Visco0}), we find a crossover scale $L_y$ at which $\nu^T_2$ in Eq. (\ref{Visco2}) becomes larger than $\nu^T_0$: $L_y \ll \A^{-2} B_0^{3} \nu^{-1}$.

Similarly, the leading order contribution to the electromotive force is found to be proportional to $B_0$ and, consequently, the $\alpha$ effect to leading order is independent of $B_0$. It is the kinematic $\alpha$ effect which can be computed by ignoring the back-reaction of the magnetic field on the flow. In the strong-shear limit ($\xi \ll 1$) this gives:
\EQA
\nonumber
\alpha_1 &\sim& - \frac{\tau_f}{(2 \pi)^3 \A^2} \int d^3 k \; k^2  H(k) \T({\bf k}) \Gamma(\frac{1}{3}) \frac{\xi^{-1/3}}{3} \\ \label{alpha0}
&\sim& - \xi^{5/3} \frac{h_0}{\nu k^2} \; . 
\ENA
Therefore, to leading order, the $\alpha$ effect is reduced by shear proportionally to $\A^{-5/3}$. This result shows, for the first time, that the $\alpha$ effect can significantly be reduced by a strong shear. In contrast to the turbulent viscosity, the $\alpha$ effect is proportional only to the non-reflectionally symmetric part of the forcing. This agrees with the expectation that the $\alpha$ effect is present only for flow with helicity, which results from the helical forcing with helicity spectrum $H$ in our case. 


To next order, we compute the correction to the $\alpha$ effect due to magnetic field. In the strong shear limit ($\xi \ll 1$) we can then obtain the following:
\EQ
\alpha_3 \sim \frac{\tau_f B_0^2}{(2 \pi)^3 \A^4} \int d^3 k \; k_y^2 k^2  H(k) \T({\bf k}) \frac{1}{3 \xi} \sim \gamma^2 \xi \frac{h_0}{\nu k^2} \; . 
\EN
This correction has obviously the opposite sign to the leading order term (\ref{alpha0}). Therefore, both the magnetic field and the shear quench the $\alpha$ effect. The ratio of these two terms is of the same order of magnitude as the ratio of the first correction to the leading order term in the turbulent viscosity. In other words, the $\alpha$ effect is suppressed by magnetic fields when $B_0 \sim \A / k R_m^{1/3}$, where $R_m = \A / \eta k^2$ is the magnetic Reynolds number. This is an important result showing that a sufficiently strong shear can modify the efficiency of the dynamo saturation, leading to a completely new scaling. This $\alpha$ quenching due to flow shear has not been investigated in previous works which did not incorporate the effect of shear or incorporated it only perturbatively suggesting the critical magnetic field to be either independent of $R_m$ \citep{Blackman96} or to scale as $R_m^{-1}$ \citep{Cattaneo96}. 

\noindent{\it Strong magnetic field.}-- In the case of a strong magnetic field ($\gamma \gg 1$), we obtain a WKB solution for the small-scale fields and then compute the turbulent transport coefficients with the help of Eq. (\ref{Forcing}). First, the turbulent viscosity is given by:
\EQA
\label{ViscoStrong}
\nu^T &=& \frac{\tau_f}{(2 \pi)^3} \int d^3 k \; E(k)  \frac{k_H^2}{k^2} \times \\ \nonumber
&& \quad \left\{ \frac{k_x^2}{4 B_0^2 k_y^2 k^2} + \frac{k_z^2 k_y^2  B_0^2}{k^2 [B_0^2 k_y^2 + \nu^2 k^4]^2} \right \} \; ,
\ENA
to leading order in $\gamma$. Assuming isotropic forcing and performing the integration on the angular variables, we find that, for strong magnetic field ($B_* \gg 1$, where $B_*= \vert B_0 \vert / \nu k$), the turbulent viscosity is reduced by $B_0$ as $B_*^{-1}$ due to alfvenization of turbulence \citep{Kim06}. Note that in 2D ($k_z = 0$) $\nu^T$ is positive, in agreement with \cite{Kim06}. Furthermore, the second term in Eq. (\ref{ViscoStrong}) is also positive. Therefore, $\nu^T$ in 3D MHD is positive with larger value compared to the 2D MHD case.

Similarly, the electromotive force can be computed to leading order, with the following result for the $\alpha$ effect:
\EQ
\label{alpha0bis}
\alpha_0 = - \frac{\tau_f}{2 (2\pi)^3} \int d^3 k \; k_y^2 \frac{H(k)}{B_0^2 k_y^2 + \nu^2 k^4} \; .
\EN
The integration of Eq. (\ref{alpha0bis}) over the angular variables gives:
\EQ
\label{BS4}
\alpha_0 = - \frac{\tau_f}{(2\pi)^2} \int_0^\infty d k \; \frac{H(k)}{\nu^2 B_*^2} \left[1-\frac{\arctan(B_*)}{B_*} \right]  \sim - B_*^{-2} \frac{h_0}{\nu k^2}\; .
\EN
Therefore, for strong magnetic field ($B_* \gg 1$) the $\alpha$ effect scales as $B_*^{-2}$. For a forcing with finite correlation time a different scaling, $B_*^{-n}$ with $2 \leq n \leq 3$, can be expected \cite{Rudiger74,Inertiel}.

While the formula (\ref{alpha0bis}) shows that in the strong magnetic regime the main quenching of the $\alpha$ effect is due to the magnetic field, shear flow is found to have an interesting effect on $\alpha$ at second order in $\gamma^{-1}$ as follows:
\EQA
\label{BS3}
\alpha_2 &=& \frac{\tau_f}{2 (2\pi)^2} \int_0^\infty d k \; \frac{H(k)}{\nu^2 B_*^4} \left(\frac{\A}{\nu k^2}\right)^2 \times \\ \nonumber
&& \qquad \left[1-\frac{2 \arctan(B_*)}{B_*} \right] \sim B_*^{-4} \xi^{-2} \frac{h_0}{\nu k^2} \; .
\ENA
For strong magnetic field ($B_* \gg 1$) we see in (\ref{BS3}) that the first correction is of the opposite sign to the leading order and that it scales as $\A^2 B_*^{-4}$. Comparing (\ref{BS3}) with (\ref{BS4}) we can easily see that the correction (\ref{BS4}) is always negligible compared to the leading order term (\ref{BS3}) as $B_0 k / \A \gg 1$ in the strong magnetic field regime. Consequently, the $\alpha$ effect is suppressed when $B_* \gg 1$, i.e.  for $B_0 \gg \Omega / k R_m$.

\noindent{\it Discussion.}-- We show an $\alpha$-quenching due to shear ($\alpha \propto \Omega^{-n}$) for the first time. Depending on the ratio of the large-scale magnetic field to the shear, we found that the $\alpha$ effect is quenched by large-scale magnetic field with different scalings with $R_m$. Specifically, the critical magnetic field strength $B_c$, above which the $\alpha$ effect is suppressed by magnetic field, is $B_c \sim \Omega / k R_m^{1/3}$ for $\gamma=B_0 k/ \A \gg 1$ while $B_c \sim \Omega / k R_m$ for $B_0 k \ll \A$, where $k^{-1}$ is the typical small scale of turbulence. For instance, in the solar tachocline ($\A \sim 3 \times 10^{-6} \, \mbox{s}^{-1}$, $\nu \sim 10^{-2} \, \mbox{m}^2 \cdot \mbox{s}^{-1}$ and $B_0 \sim 6 \, \mbox{m} \cdot \mbox{s}^{-1}$), the strong magnetic field regime ($B_0 k \gg \A$) is valid on scales less than $2 \times 10^{6} \; \mbox{m}$, which is approximately the size of the tachocline. Therefore, not only magnetic field but also shear can dramatically quench the $\alpha$ effect hindering large-scale dynamos.

The turbulent viscosity is reduced by strong magnetic field (with a scaling $B_*^{-2}$) while in the opposite limit of weak magnetic field is quenched by shear (with a scaling $\A^{-2}$). In both cases, the turbulent viscosity is positive with a larger value in 3D than in 2D. In the weak magnetic field case, magnetic field becomes important on scale $L_y < \A^{-2} B_0^{3} \nu^{-1}$. In  case of the Sun, $L_y < 5.4 \times 10^{15} \, \mbox{m}$, suggesting that turbulent viscosity is severely quenched by magnetic field. 

These results have crucial implications for dynamics and angular momentum transport in astrophysical or laboratory plasmas which are often envisioned to be efficient. In particular,  quenching by shear and/or magnetic field should be incorporated when assessing the efficiency of dynamo, e.g. the dynamo number $D$. For instance, our result $\alpha \propto \A^{-5/3}$ makes $D \propto \A^{-2/3}$, decreasing with strong shear rather than increasing ($D \propto \A$) as previously thought. Furthermore, the nonlinear dependence of turbulent viscosity on $B_0$ can offer an interesting mechanism for the time variability such as torsional oscillation in the Sun \cite{Rudiger90}. It will be interesting to extend our theory to incorporate the effects of rotation which will consistently give rise to $\alpha$ effect and non-diffusive momentum transport, ($\Lambda$ effect), due to shear-induced anisotropy \cite{RotShearAA}. How the $\Lambda$ effect, $\alpha$ effect, turbulent viscosity and particle transport are affected by rotation, magnetic field and shear would be of great interest with important implications. The controversial issue on the quenching of $\beta$ effect in 3D MHD should also be investigated by incorporating the gradient of large-scale magnetic field. These issues will be addressed in future publications.

\begin{acknowledgments}
This work was supported by U.K. PPARC Grant No. PP/B501512/1.
\end{acknowledgments}


\end{document}